\documentclass{ws-procs975x65}

\def\l{\ell}
\def\lm{\ell m}

\begin{document}

\title{BINARY BLACK HOLE MERGER WAVEFORMS IN THE
EXTREME MASS RATIO LIMIT}

\author{THIBAULT DAMOUR}
\address{Institut des Hautes Etudes Scientifiques, 35 route de Chartres, 91440 Bures-sur-Yvette, France}
\author{ALESSANDRO NAGAR}
\address{Dipartimento di Fisica, Politecnico di Torino, and INFN, sezione di Torino, Torino, Italy}

\begin{abstract}
  We discuss the   transition   from  quasi-circular  inspiral  to plunge of a
  system of two nonrotating black holes of masses $m_1$  and  $m_2$ in the
  extreme mass ratio limit $m_1m_2\ll (m_1+m_2)^2$. In this limit,
  we compare the merger waveforms obtained by two different methods:
  a {\it numerical} (Regge-Wheeler-Zerilli) one, and an {\it analytical}
(Effective One Body) one. This is viewed as
  a contribution to the matching between analytical and numerical methods.
\end{abstract}

\bodymatter

\section{Introduction}
\label{intro}

The last months have witnessed a decisive advance in Numerical Relativity,
with different groups being able to simulate the merger of two black holes 
of comparable masses~\cite{num}.
Since such binary black holes systems (of a total mass$\sim30 M_{\odot}$) 
are believed to be among the most promising sources of gravitational waves for 
the ground based detectors like LIGO and VIRGO, this breakthrough 
raises the hope to have, for the first time, a reliable estimate of the 
complete waveform by joining together Post-Newtonian (PN) 
and Numerical Relativity results. We recall that PN techniques have provided 
us with high-order results for describing the motion
and radiation~\cite{PN} of binary systems, and that further
techniques have been proposed for {\it resumming} the PN results~\cite{DIS,BD99}, 
thereby allowing an analytical description of the gravitational waveform 
emitted during the transition between inspiral and plunge, and even during 
the subsequent merger and ringdown phases. We now face the important task 
of constructing accurate {\it complete waveforms} by {\it matching} together 
the information contained in {\it Post-Newtonian} and {\it Numerical Relativity} 
results. We view the present work as a contribution towards this goal
(see also Refs.~\cite{BCP06a,Baker06b}).
\begin{figure}[t]
\begin{center}
\includegraphics[width=5.50 cm,height=5.00 cm]{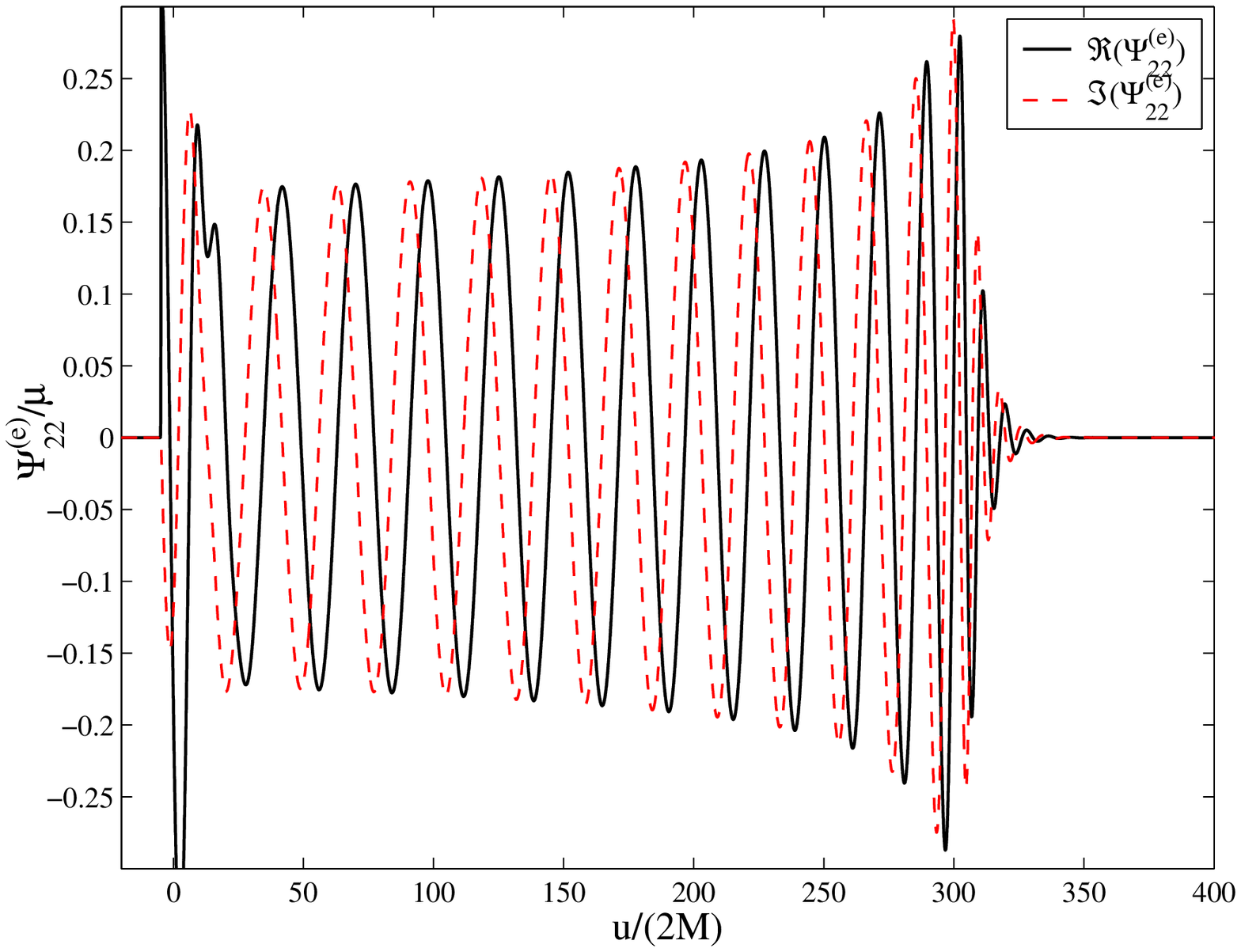}\qquad
\includegraphics[width=5.50 cm, height=5.00 cm]{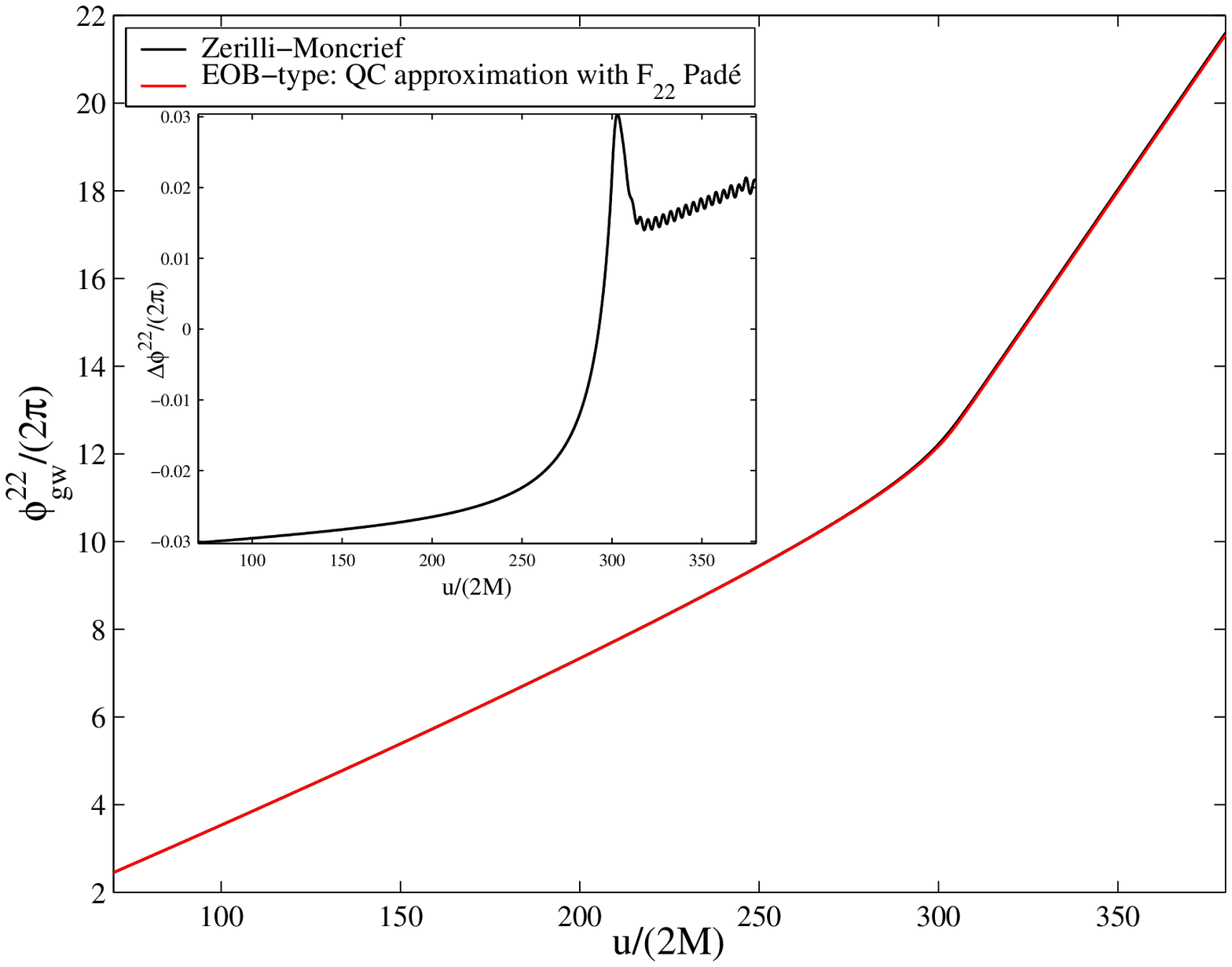}
\caption{\label{fig1} {\it Left panel: }Gravitational waveforms (real and imaginary part) for $\l=m=2$
generated by a plunge from an initial separation $r=7M$. {\it Right panel:} Comparison between the corresponding 
gravitational-wave phase and that obtained from analytically matching a 3PN improved quadrupole-type 
formula to a superposition of quasi-normal modes.}
\end{center}
\end{figure}

The present work belongs to a scientific lineage which was
started by Regge and Wheeler~\cite{RW57}, Zerilli~\cite{zerilli70}, 
Davis, Ruffini, Press and Price~\cite{DRPP}  and Davis, Ruffini and 
Tiomno~\cite{DRT}. Refs.~\cite{DRPP,DRT} studied  the gravitational 
wave emission due to the {\it radial plunge}
(from infinity) of a particle into a Schwarzschild black hole, as a 
model for the head-on collision of two black holes in the extreme mass
ratio limit.
Here we consider, for the first time, the {\it transition}
from the quasi-circular adiabatic inspiral phase to the plunge phase 
in extreme-mass-ratio binary black hole systems. We shall be able to
do that by getting round a limitation of
the original Regge-Wheeler-Zerilli test-particle approach: that
of requiring the test particle to follow an exact {\it geodesic}
of the Schwarzschild background.
We bypass this stumbling block by using an improved form of PN theory:
the Effective One Body (EOB) approach to the general
relativistic two body dynamics. This approach has been recently
proposed to
study the transition from inspiral to plunge in the {\it comparable-mass}
case~\cite{BD99}.
It describes the dynamics of
a binary system in terms of two separate ingredients: (i) a Hamiltonian
$H_{\rm EOB}(M,\mu)$ describing the {\it conservative} part of the 
relative dynamics, and (ii) a non-Hamiltonian supplementary force 
${\cal F}_{\rm EOB}(M,\mu)$ approximately describing the reaction
to the loss of energy and angular momentum along quasi-circular 
orbits. [Here, $M \equiv m_1 + m_2$, and
$\mu = m_1 m_2/M$.]
The badly convergent PN-Taylor series giving the angular
momentum flux is {\it resummed} by means of Pad\'e 
approximants~\cite{DIS}.

\section{Results}
\label{sec:results}
We summarize here our tools and main results (see Refs.~\cite{NDT06,DN06} for more information).
We describe, in the extreme mass ratio limit, the relative dynamics of the binary 
system of two non-rotating black holes by that of a ``particle'' (of mass $\mu\ll M$) 
that follows a non-geodesic inspiral driven by radiation-reaction
(until a quasi-geodesic  plunge) on a  (quasi-)Schwarzschild background 
of mass $M$. We then use the multipolar
Regge-Wheeler-Zerilli
perturbation theory around such a black hole to compute
the gravitational wave emission in the approximation $\mu\ll M$.
Einstein equations then lead to two (even/odd) {\it decoupled} wave-like
equations (with source terms $S^{(\rm e/o)}_{\lm} \propto \mu$)
for two (even/odd) master functions $\Psi^{(\rm e/o)}_{\lm}$~\cite{NR05}.

Fig.~1 shows the $\l=m=2$ waveform  generated
by a binary system with initial relative separation $r=7M$ 
\footnote{Note that the other multipoles give a non-neglible contribution, due to 
the strong asymmetry of the system.}. After a ``chirp-like'' increase of the frequency 
and modulus during the inspiral, the latter reaches a maximum when the particle
crosses the light ring ($r=3M$) and eventually the waveform decays in a
quasi-normal mode ringing phase.

The so-constructed ``exact'' {\it numerical}
waveforms are then compared with  semi-{\it analytical} ones,
constructed within the EOB framework
and philosophy. We recall that the basic idea of the EOB framework is to 
produce quasi-analytical waveforms by patching together a quadrupole-type
waveform during the inspiral and plunge to a QNM-type waveform 
after merger. We give an example of this numerical-analytical comparison 
in the right panel of Fig.~\ref{fig1}. The gravitational wave phases $\phi^{22}_{\rm gw}$ 
obtained by two different methods, one numerical and the other semi-analytical,
are compared. In the first method, $\phi^{22}_{\rm gw}$ is given from time-integration of
the instantaneous gravitational wave frequency obtained as 
$M\omega_{\rm gw}^{22}=-\Im(\dot{\Psi}^{(\rm e)}_{22}/\Psi^{(\rm e)}_{22})$.
The second method computes an approximate waveform in the following
way: before crossing the light ring one uses a (Pad\'e resummed) 3PN-improved
quadrupole-type formula to compute the waveform from the EOB dynamics.
After crossing the light ring, the previous quadrupole-type signal, taken in
the quasi-circular (QC) approximation, is {\it matched} to a superposition of 
the first five QNMs of the black hole. Then one computes the phase of the 
matched analytical waveform by integrating the corresponding instantaneous 
gravitational wave frequency. The maximum difference between the ``exact'' phase and the
``effective one body'' phase turns out to be less than $3\%$ of a cycle (see inset).

\vfill

\end{document}